\documentclass[twoside,a4paper,11pt]{sea10}
\usepackage{graphicx}
\usepackage{hyperref}
\usepackage{movie15}
\begin{document}
\pagenumbering{arabic}
\pagestyle{myheadings}
\thispagestyle{empty}
{\flushleft\includegraphics[width=\textwidth,bb=58 650 590 680]{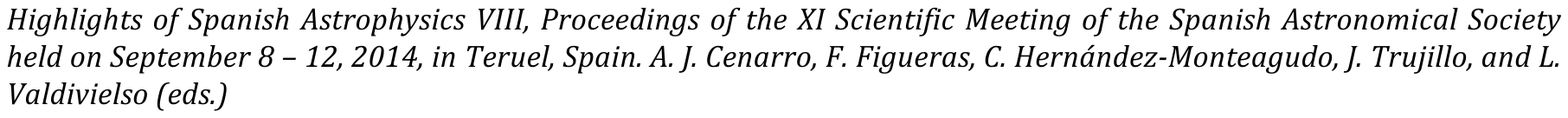}}
\vspace*{0.2cm}
\begin{flushleft}
{\bf {\LARGE
%
Adaptive Optics and Lucky Imager (AOLI): presentation and first light.
%
}\\
\vspace*{1cm}
%
Velasco, S.$^{1,2}$,
Rebolo, R.$^{1,2,3}$,
Mackay, C.$^{4}$,
Oscoz, A.$^{1,2}$,
King, D.L.$^{4}$,
Crass, J.$^{4}$,
D\'iaz-S\'anchez, A.$^{5}$,
Femen\'ia, B.$^{6}$,
Gonz\'alez-Escalera, V.$^{1,2}$,
Labadie, L.$^{7}$,
L\'opez, R.L.$^{1,2}$,
P\'erez Garrido, A.$^{5}$,
Puga, M.$^{1,2}$,
Rodr\'iguez-Ramos, L.F.$^{1,2}$
and
Zuther, J.$^7$
%
}\\
\vspace*{0.5cm}
%
$^{1}$Instituto de Astrof\'isica de Canarias, c/V\'ia L\'actea s/n, La Laguna, E-38205, Spain\\
$^{2}$Departamento de Astrof\'isica, Universidad de La Laguna, La Laguna, E-38200, Spain\\
$^{3}$Consejo Superior de Investigaciones Cient\'ificas, Madrid, Spain\\
$^{4}$Institute of Astronomy, University of Cambridge, Madingley Road, Cambridge CB3 0HA, United Kingdom\\
$^{5}$Universidad Polit\'ecnica de Cartagena, Campus Muralla del Mar, Cartagena, Murcia E-30202, Spain\\
$^{6}$W. M. Keck Observatory, 65-1120 Mamalahoa Hwy., Kamuela, HI 96743, Hawaii, USA\\
$^{7}$I. Physikalsiches Institut, Universit\"{a}t zu K\"{o}ln, Zülpicher Strasse 77, 50937 K\"{o}ln, Germany

%
\end{flushleft}
%
\markboth{
Adaptive Optics and Lucky Imager (AOLI): presentation and first light.
}{ 
%
Velasco, S. et al.
%
}
\thispagestyle{empty}
\vspace*{0.4cm}
\begin{minipage}[l]{0.09\textwidth}
\ 
\end{minipage}
\begin{minipage}[r]{0.9\textwidth}
\vspace{1cm}
\section*{Abstract}{\small
%

In this paper we present the Adaptive Optics Lucky Imager (AOLI), a state-of-the-art instrument which makes use of two well proved techniques for extremely high spatial resolution with ground-based telescopes: Lucky Imaging (LI) and Adaptive Optics (AO).

AOLI comprises an AO system, including a low order non-linear curvature wavefront sensor together with a 241 actuators deformable mirror, a science array of four 1024x1024 EMCCDs, allowing a 120x120 down to 36x36 arcseconds field of view, a calibration subsystem and a powerful LI software. Thanks to the revolutionary WFS, AOLI shall have the capability of using faint reference stars ({\it I\/} $\sim$ 16.5-17.5), enabling it to be used over a much wider part of the sky than with common Shack-Hartmann AO systems. 

This instrument saw first light in September 2013 at William Herschel Telescope. Although the instrument was not complete, these commissioning demonstrated its feasibility, obtaining a FWHM for the best PSF of 0.151$\pm$0.005 arcsec and a plate scale of 55.0$\pm$0.3 mas/pixel. Those observations served us to prove some characteristics of the interesting multiple T Tauri system LkH$\alpha$ 262-263, finding it to be gravitationally bounded. This interesting multiple system mixes the presence of proto-planetary discs, one proved to be double, and the first-time optically resolved pair LkH$\alpha$ 263AB (0.42 arcsec separation).

\normalsize}
\end{minipage}
%
%
%
\section{Introduction \label{intro}}
Adaptive Optics (AO) is the main technique to improve the spatial resolution of large ground-based telescopes, see \cite{1993ARA&A..31...13B}. It has turned to provide excellent results on offering diffraction limited images in the near infrared (NIR) due to the minor effects of turbulence in this range. Unfortunately, by now there is a lack of efficient AO systems in the optical bands (see \cite{2013ApJ...774...94C} and \cite{2005ApJ...629..592G}) except for solar telescopes, as in \cite{2012AN....333..863B}. Besides, AO systems on duty today are able to achieve high success in correcting the incoming wavefront errors at high degrees by using Shack-Hartmann (SH) sensors, but they requiere very bright, and hence scarce, reference stars or a laser non-natural star.

The Lucky Imaging (LI) technique, as suggested by \cite{1964JOSA...54...52H} and named by \cite{1978JOSA...68.1651F}, was born as an alternative to AO to reach the diffraction limit in the optical bands. Images are taken at a very high speed in order to sample those intervals in which the atmosphere inside the collector tube through which the wavefront travels can be regarded as stable. If the best fraction of a bunch of images, those with smaller Strehl pattern, are stacked in a shift-and-add process, the equivalent to a near-diffraction limit observation is obtained, see \cite{2006A&A...446..739L}. The fraction of images selected for each target depends on the atmospheric conditions, see \cite{1981JOSA...71.1138B}. Examples of LI instruments are LuckyCam \cite{2002A&A...387L..21T} and FastCam \cite{2008SPIE.7014E..47O}, both of them have set the basis for AOLI.

In the next section we briefly present AOLI. In section \ref{sec:veri}, we describe its first light observations.

\section{Description of the instrument}

\begin{figure}
\center
\includegraphics[width=8cm]{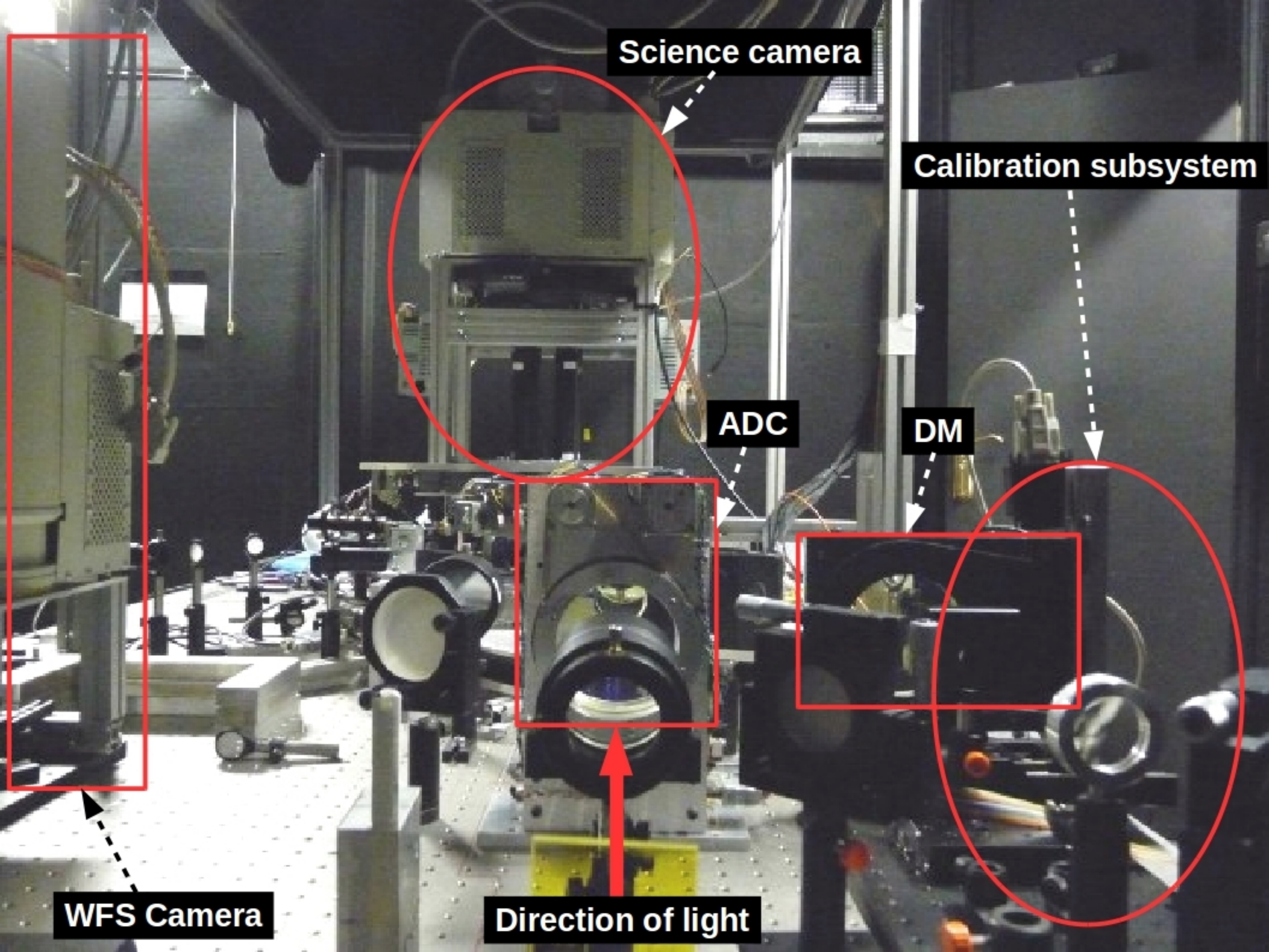} 
\caption{\label{fig:aolifoto} AOLI mounted in the Naysmith GHRIL focus at WHT in September, 2013, as seen from the telescope.
}
\end{figure}

AOLI (figure \ref{fig:aolifoto}) is being developed by a consortium composed by the IAC, the IoA, K\"{o}ln University (KU) and UPCT. It is conceived as a visitor instrument at the WHT's Nasmyth platform. It consists of seven major subsystems: front-end, mechanical support, calibration unit, science camera, AO unit, control software and data analysis. Complementary information and optical layouts can be found in \cite{2014SPIE.9147E..1TM} and \cite{2012SPIE.8446E..21M}. 
 
The collected light, after passing through the Canary de-rotator and a collimator lens, goes through an Atmospheric Dispersion Corrector (ADC) which allows AOLI to work away from zenith without significant distortions. Then, and after being reflected at the deformable mirror (DM) and a flat mirror, the beam is focused with a camera lens into the pickoff mirror. In there, light coming from the reference star is deflected towards the wavefront sensor (WFS) in where a pupil of 2 mm is imaged. The rest of the light not deflected by the pick-off mirror is sent to the science camera which contains a set of lenses allowing a range of magnification scales giving a corresponding field of view of about 36$\times$36 arcsec up to 120$\times$120 arcsec.

\subsection{Science camera}
\begin{figure}[!h]
\centering
\includegraphics[width=7cm]{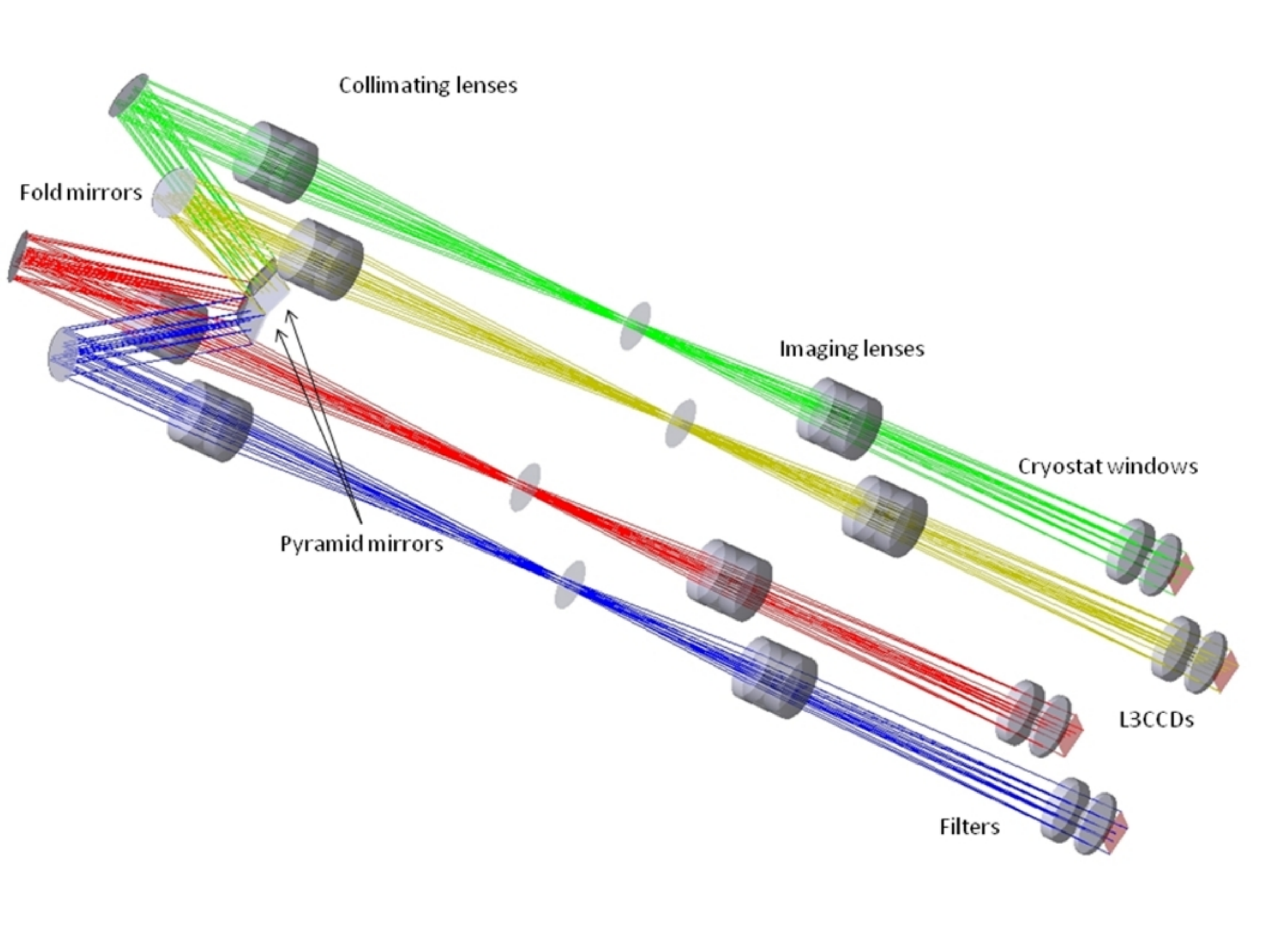}
\caption{AOLI's science camera layout.}
\label{fig:camara_layout}
\end{figure}

A pyramid mirror architecture is placed at the entrance of the camera system \ref{fig:camara_layout} to re-image contiguous zones, each to a separate detector. Furthermore, the whole imaging structure can be moved in XYZ axis for alignment, center and focusing purposes. The wide-field science camera is optimised for the 500 nm to 1 micron wavelength range with an array of four photon counting, electron multiplying, back illuminated E2V EMCCD detectors. The four 1024x1024 pixels detectors, with very high quantum efficiency (peak over 95 per cent), and our own control electronics are stacked into a cryostat to improve their performance and with its own filter wheel each. This allows the use of a narrowband filter for the science object with a broad band filter for the reference star. These EMCCDs are able to offer images at a 30 MHz pixel rate, meaning 25 frames per second (fps). 

\subsection{AO Subsystem}
\begin{figure}[!h]
\centering
\includegraphics[width=0.7\linewidth]{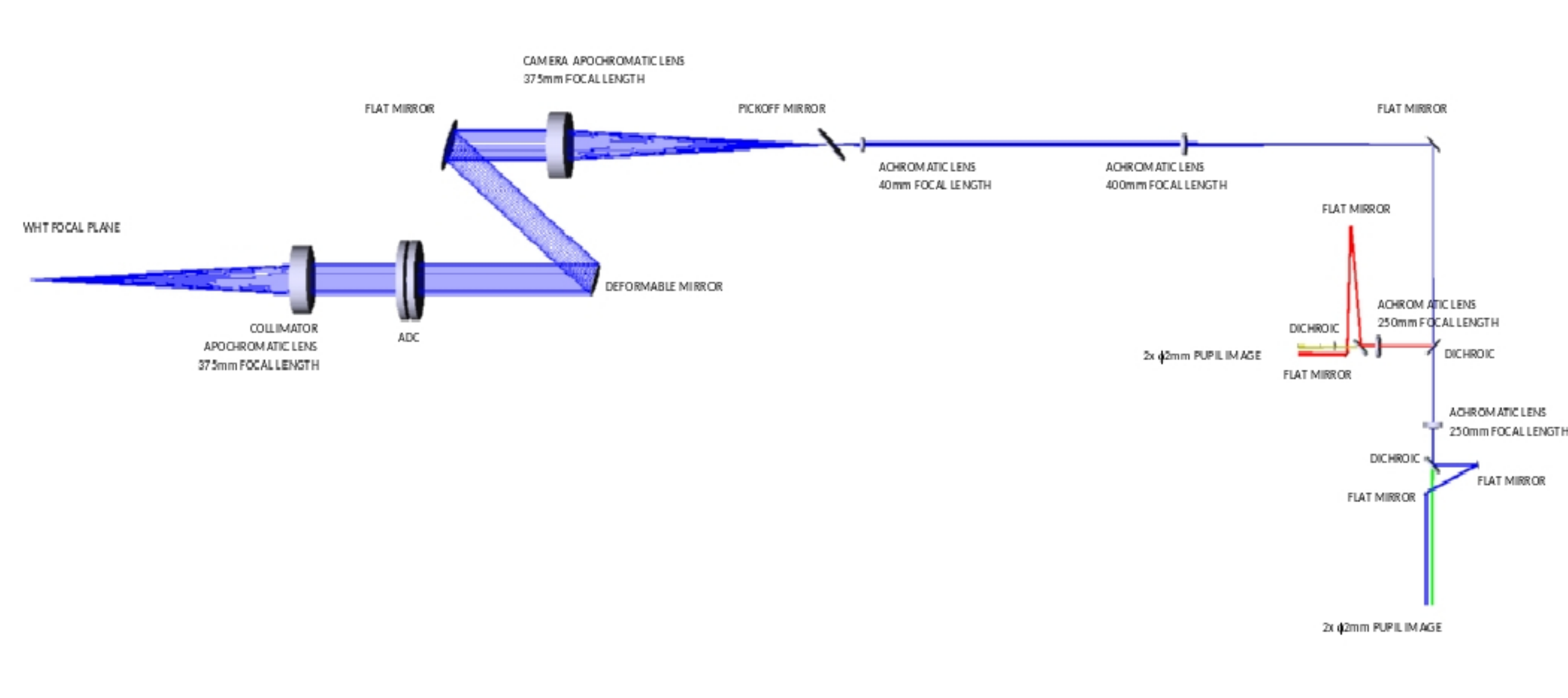}
\caption{AOLI's WFS layout.}
\label{fig:ao_layout}
\end{figure}

The AO subsystem \ref{fig:ao_layout} consists of a low order curvature WFS with two photon counting EMCCD detectors, the same ones than in the science unit but at higher frame rates (around 100 fps) reached by reducing the read out format to 1024$\times$256 pixels. Optical path details and more layouts can be found in \cite{2012SPIE.8447E..0TC}. 

Two curvature WFS are being developed for AOLI: a non-linear (\cite{2011SPIE.8149E..09M} and \cite{2007SPIE.6691E..0GG}) and a geometric one \cite{2013OptEn..52e6601F}. Curvature sensors are typically ten times more sensitive than Shack-Hartmann (SH) WFS for the same degree of correction, especially for the low-order errors due to turbulence \cite{2006PASP..118.1066R}. AOLI is designed to perform low-order corrections with much fainter stars ({\it I\/} $\sim$ 16.5-17.5), enabling the instrument to be used over a much wider part of the sky than with SH WFS. A final decision on which sensor will be finally installed at AOLI will be taken on the basis of their results.

\subsection{Other subsystems}

A calibration unit to test AOLI's performance at lab in the conditions that the instrument would likely have at telescope is included, see \cite{2014SPIE.9147E..7VP}. This subsystem comprises some phase plates to emulate local atmospherical disturbances reproducing different turbulent layers that allow to test AOLI's WFS and pupil reconstruction.

The unique acquisition software developed takes AOLI to a priority position. It brings the possibility of processing the speckle image on real time, even allowing the observer to check a live-view of it during the observation night. It not only becomes AOLI into a powerful scientific instrument but also reduces in a significant way the amount of data to be stored by computers.

\section{Science verification}
\label{sec:veri}

On September 24th and 25th 2013, AOLI was installed at WHT's Nasmyth platform for a first commissioning to test all AOLI subsystems and its overall efficiency without a fully developed AO subsystem. Although the observations were carried out under quite adverse weather conditions and with only one of the four science EMCCDs, it was possible to obtain several thousand images of scientific interest.

\subsection{Plate scale}
\label{sec:plate}

Binary stars and globular clusters were observed to test the optical alignment and to stablish a useful calibration scale using on-sky targets. The particular characteristics of Globular Clusters (GCs) -presence of thousands of stars with different spectral type and great dynamic range- make them isolated laboratories in where to test and calibrate new LI instruments. We selected M15 as it is a compact GC with a dense core that has been previously astrometrized using {\it HST\/} data, see \cite{2014arXiv1410.5820B} and \cite{2002AJ....124.3255V}. Moreover, this GC has been extensively studied by our group using LI \cite{2012MNRAS.423.2260D}.

We selected the best 10 per cent out of 1029 images of the core of M15 from 75 seconds of on-source observations at WHT. To calculate the plate scale we have cross-correlated them with the catalogues by \cite{1994AJ....107.1745Y} and \cite{2002AJ....124.3255V}, measuring a plate scale of 55.0$\pm$0.3 mas/pixel.

\subsection{Observations of the LkH$\alpha$ 262-263 T Tauri system}

LkH$\alpha$ 263 is a triple T Tauri system inside the MBM12 cloud at  275 pc from the Sun \cite{2001ApJ...560..287L}. It includes two main bright non-spectroscopically resolved M2-M4 (see \cite{2009A&A...497..379M} and \cite{2001ApJ...560..287L}) components, A and B, separated by 0.4 arcsec, and a third fainter C component 4 arcsec away. This C component, an optically thick edge-on disk hosting an M0 star, was discovered with adaptive optics in the near-IR band by \cite{2002ApJ...571L..51J}. LkH$\alpha$ 262 is another T-Tauri M0 \cite{2001ApJ...550L.197J} star in the proximity, 15.5 arcsec away, of LkH$\alpha$ 263.

\begin{figure}[!hr]
\center
\includegraphics[width=6cm]{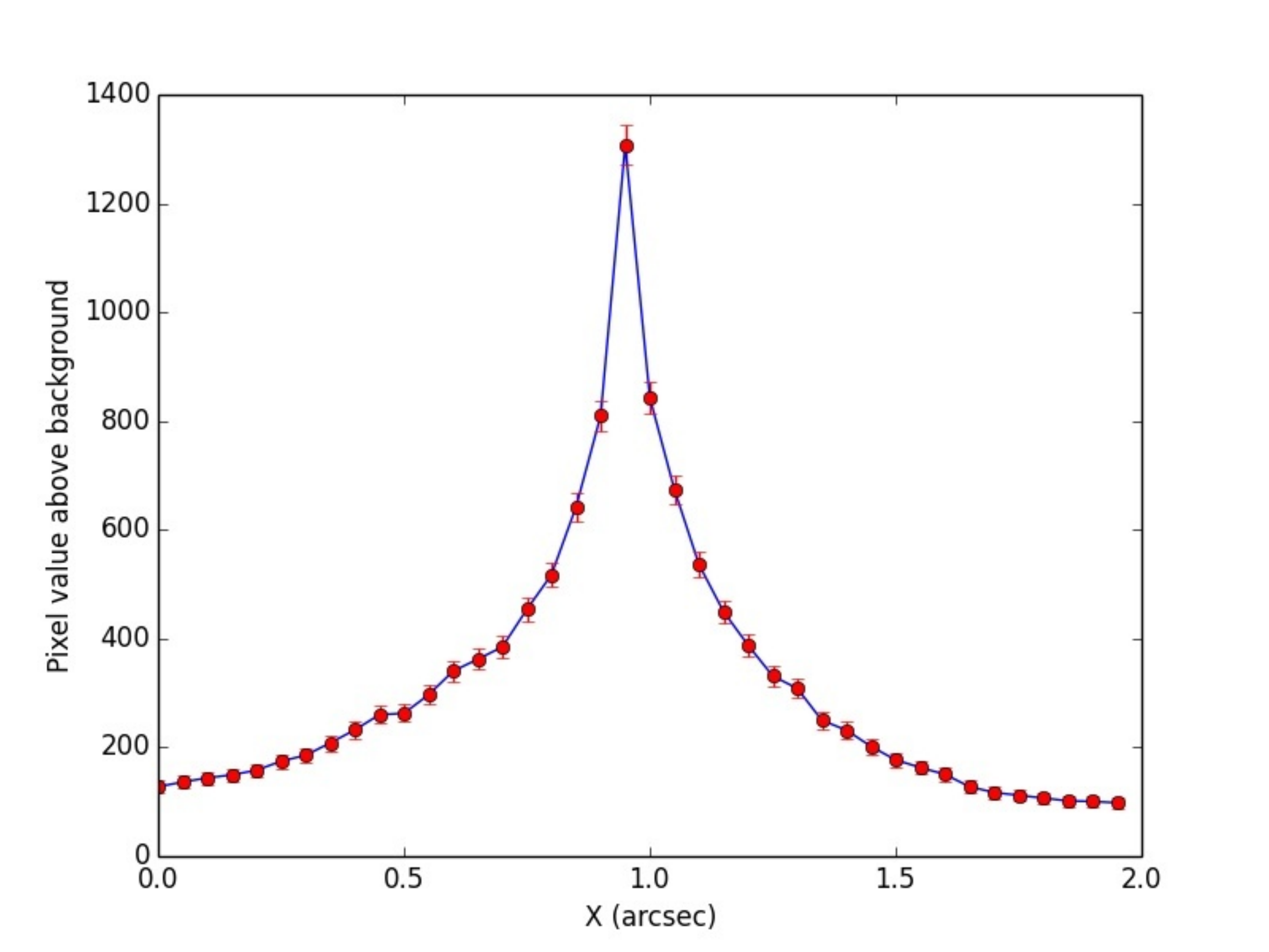} 
\caption{\label{fig:stf} AOLI's PSF for LkH$\alpha$ 262 in the {\it i'\/} band.
}
\end{figure}

The LkH$\alpha$ 262-263 T Tauri system was observed in the night of 2013 September 24th with AOLI in the standard  {\it i'\/}  band [769.5 nm central wavelength, 137 nm full width at half-maximum (FWHM)] at WHT. Both targets were included within the FOV of the first 1024x1024 EMCCD camera detector. The lowest magnification was used and the plate scale was estimated to be 55.0$\pm$0.3 mas/pixel. Despite poor seeing conditions we could obtain a FWHM for the best PSF of 0.151$\pm$0.005 arcsec, see figure \ref{fig:stf}.

In our optical observations we have noticed that: a) There is orbital motion of the pair LkH$\alpha$ 262 AB. b) Component LkH$\alpha$ 263 C is comoving with the AB pair. c) LkH$\alpha$ 262 and 263 are also comoving. d) The likely existence of a close companion to LkH$\alpha$ 262. e) Besides the resolved edge-on thick disk around LkH$\alpha$ 263 C, the SEDs show the presence of disks also around either 263 A, 263 B or both of them and around LkH$\alpha$ 262. More details are given in \cite{Velascoposter} and \cite{Velasco}

%
\small  
%
%

%

%
\end{document}